# Influence of nonstoichiometry and iron doping on optical absorption in $KNbO_3$ single crystal

M. Milata, K. Wójcik and K. Zieleniec

Institute of Physics, University of Silesia,
ul. Uniwersytecka 4, PL-40-007 Katowice, Poland

## 1. Introduction

Ferroelectric potassium niobate ($KNbO_3$) is a typical representative of the perovskite crystals family. It exhibits the same sequence of structural phases as $BaTiO_3$, but at higher temperatures. $KNbO_3$ single crystals are well known as material with large nonlinear optical and electro–optical coefficients, which make them attractive for applications in nonlinear optic as laser light modulators or the second harmonic generators. Lattice defects exert an important influence on the physical properties of the crystals. Sufficiently high concentration of defects, associated with nonstoichiometry and/or doping may cause colouration of crystals and increase in their electric conductivity that makes technical application of these crystals difficult or even impossible. Nonstoichiometry and doping may cause generation both point defects and extended defects known as shear structures or Ruddlesden - Popper phases [1,2,3,4,5]. The aim of these studies was to examine the optical absorption spectra for $KNbO_3$ single crystals grown from the solution $Nb_2O_5$– $K_2CO_3$. The samples with different degree of oxygen defectation and also crystals doped with iron were used for investigations.

## 2. Experimental

### *2.1 Crystal growth*

$KNbO_3$ single crystals (labelled as A, B, C) were obtained from the 0,565 mole $K_2CO_3$– 0,435 mole $Nb_2O_5$ melted solution [1, 6]. The composition of the solution and the temperature range of the crystal growth were determined from the $K_2O$–$Nb_2O_5$ phase diagram given by Reisman and Holtzberg [7].

The crystals doped with iron ($KNbO_3$: Fe labelled as D) were obtained from the solution: $0.565K_2CO_3$–$0.434Nb_2O_5$–$0.0001Fe_2O_3$. For crystal growth the following reagents were used: $Nb_2O_5$ (Fluka, specpure 99.9%), $K_2CO_3$ (POCh, analar purity, 99.5%), $Fe_2O_3$ (POCh, analar purity, 99.5%).

The $K_2CO_3$–$Nb_2O_5$ mixture was homogenized in an open platinum crucible at $1050^0C$ for 12 hours. Then the temperature was lowered to $950^0C$ with the cooling rate $dT/dt = 4K \cdot h^{-1}$. After the solvent excess was poured off, crystals were cooled slowly down to room temperature in the furnace. The following chemical reaction took place in the melted solution during homogenization: $K_2CO_3 + Nb_2O_5 \rightarrow 2KNbO_3 + CO_2$. The crystal growth was performed in the presence of titanium dioxide ($TiO_2$), placed in alumina crucibles inside the furnace nearly the platinum crucible. The reaction: $2TiO_2 \rightarrow Ti_2O_3 + 1/2O_2$ should protect the demanded oxidizing atmosphere during crystal growth. Crystals obtained in this way are transparent and colourless. These crystals we labelled as A crystals. Crystals with two different degrees of oxygen reduction were obtained by homogenization at $1080^0C$ and $1100^0C$, respectively. In this way blue (labelled as B) and navy–blue (labelled as C) crystals were obtained. From $K_2CO_3$–$Nb_2O_5$–$Fe_2O_3$ mixture the Fe–doped potassium niobate crystals were obtained by homogenization at $1080^0C$. They show light grey colour and we labelled their as D crystals. Technological conditions of crystal growth are presented in Table 1.

Table 1. Technological conditions of crystal growth.

| System $0.585K_2CO_3 - 0.435Nb_2O_5$ | | cooling rate $\frac{dT}{dt} = 4\frac{K}{h}$ | |
|---|---|---|---|
| crystal | colouration | crystallization range [$^0C$] | remarks |
| A | colourless | 1050 - 950 | in the presence of $TiO_2$ |
| B | blue | 1080 - 950 | reduced |
| C | navy blue | 1100 - 950 | strongly reduced |
| D | light grey | 1080 - 950 | doped with $0.01\%mol.Fe_2O_3$ |

*2.2 Light absorption*

Optical absorption was measured in the wavelength range from 350 nm to 1250 nm using an SPM–2 mirror monochromator made by Carl Zeiss Jena. Plate shaped samples were obtained by grinding and polishing. A diamond abrasive compound was used in the polishing process. The samples were polydomain with a–type domain predominance. The optical transmission $T_R$ and the reflection $R$ were measured at room temperature. It was possible to calculate the absorption coefficient $\alpha$ from this measurement.

Using a well-known method [8] the dependence $\alpha = f(h\nu)$ was analysed and the form of the absorption edge and the width of the energy gap $E_G$ was determined. In the case of exponential absorption edge (iron doped crystal D), the approximated value of energy gap $E_G^*$ was determined from the $D = D(\lambda)$ dependence extrapolated to zero, where $D = \log\left(\dfrac{1}{T_R}\right)$ is an optical density. Results are shown in Table 2.

Table 2. Width of the energy gap in $KNbO_3$ single crystals at room temperature.

| crystal | $E_G$ [eV] direct | $E_G$ [eV] direct forbidden | $E_G$ [eV] indirect | $E_G^*$ [eV] exponential edge | remarks |
|---|---|---|---|---|---|
| A | 3.19 | 3.06 | 3,01 | - | colourless |
| B | - | 3.00 | 2.91 | - | blue |
| C | - | 2.95 | 2.87 | - | navy blue |
| D | - | - | - | 3.09 | $KNbO_3$ + 0.02 at % Fe light grey |

Results of the optical transmission measurement obtained at room temperature are shown in Figure 1. The impurity absorption band for reduced samples B and C is clearly visible. The minimum of the transmission appears at the wavelength $\lambda = 1115$ nm (1.10 eV). It is noteworthy that such absorption band is not observed for both colourless crystal A and Fe–doped crystal D.

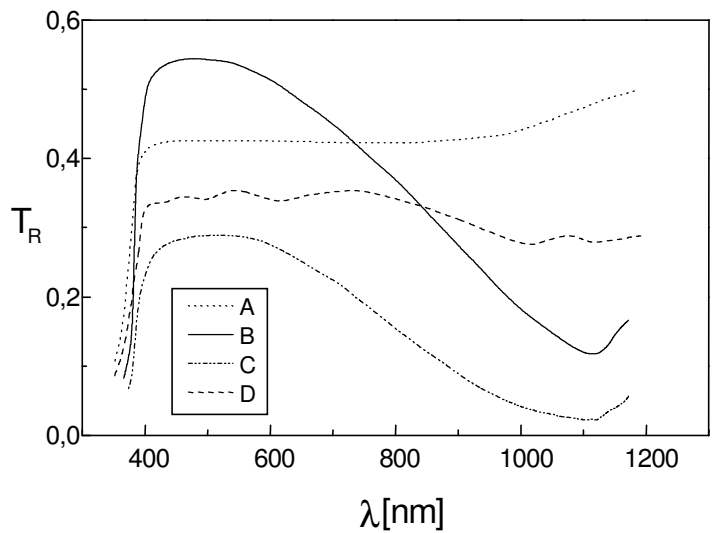

Fig. 1 Optical transmission versus wavelength at room temperature.

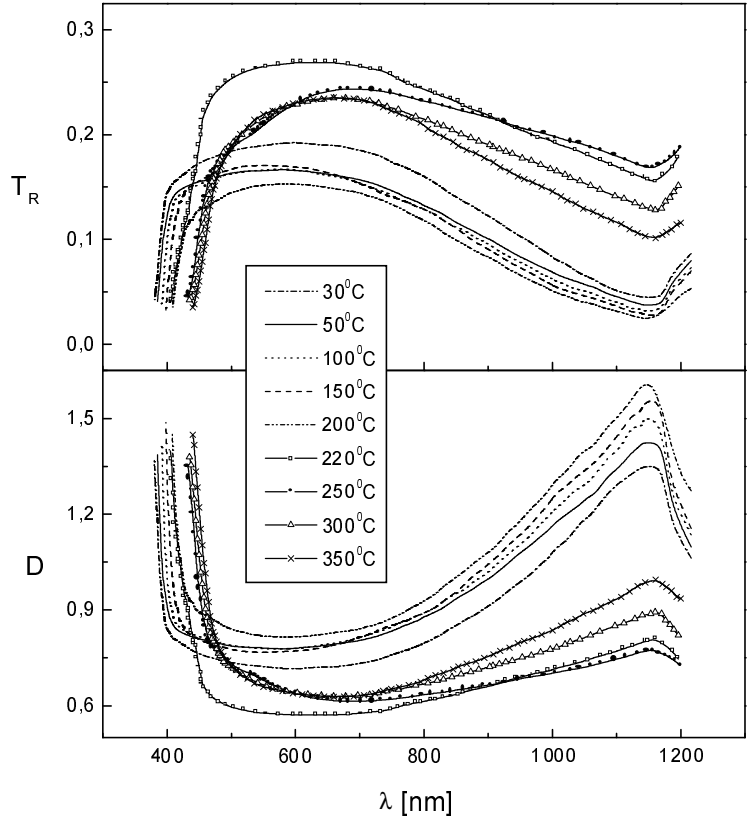

Fig. 2 Wavelength dependencies of optical transmission $T_R = T_R(\lambda)$ and optical density $D=D(\lambda)$ for reduced crystal $KNbO_3$–C at different temperatures.

Dependencies of the transmission $T_R$ and the optical density $D$ versus the wavelength $\lambda$ obtained at various temperatures are presented in Figure 2. The marked increase in the optical transmission $T_R(\lambda)$ and the shift of the absorption edge at $250^0C$ are clearly visible. These phenomena are associated with the structural phase transition from Bmm2 orthorhombic phase to P4mm tetragonal phase occurred at this temperature, that is confirmed by result of dielectric permittivity measurement (fig.3a). The impurity absorption band exists up to $350^0C$ although it vanishes gradually.

Dependencies of the dielectric permittivity and the loss tangent versus temperature for radio frequencies obtained for crystal A, are presented in Figure 3a and b respectively.

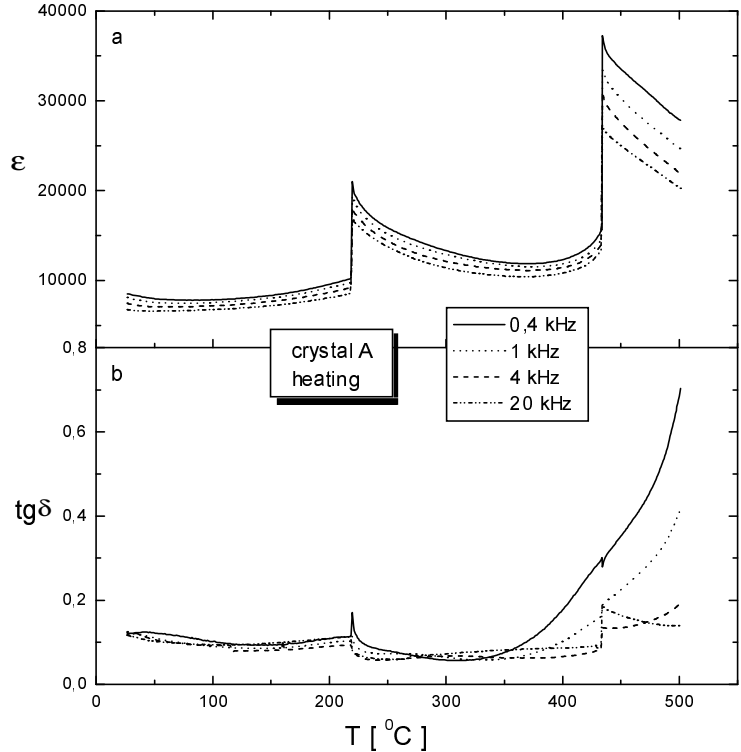

Fig. 3 Dielectric permittivity and loss tangent versus temperature.

## 3. Discussion

Both nonstoichiometry and doping exert considerable influence on physical properties of single crystals examined. Complicated phase diagram of the solution $K_2O$–$Nb_2O_5$ [7] suggests that stoichiometric deviations can appear in crystals during the growing process. The deviations from proper composition $Nb_2O_5 \cdot K_2O$ to $3K_2O \cdot Nb_2O_5$ or $2K_2O \cdot 3Nb_2O_5$ may appear in crystal as results of local temperature or composition fluctuations. Enriched or deficient in $K_2O$ regions can appear in single crystals obtained. These phenomena can lead to the creation of the layer structures known as Ruddlesden – Popper phases [3, 4, 5] or to the creation of extended defects known as shear planes [2]. Moreover, reduction processes bring about appearing of the oxygen deficiency. Oxygen deficiency leads to appearing of point defects or extended defects at the high level of the reduction.

### 3.1 Reduction of $KNbO_3$ crystals

The reduction process of stoichiometric crystals may be given as follows:

$$KNbO_3 \rightarrow KNbO_{3-y} + 1/2yO_2 \uparrow \qquad (1)$$

This process occurred during growth of crystals at high temperature ($1080^0C$–$1100^0C$) and in absence of $TiO_2$ in the furnace. Thus, $KNbO_3$ crystals (samples B and C) with oxygen deficiency are obtained and vacancies appeared in the oxygen sublattice.

The chemical composition of the crystal with the oxygen deficiency may be written as follows:

$$KNbO_{3-y} \equiv K^{1+}Nb^{5+}O^{2-}_{3-y}(yV_O) \qquad (2)$$

Oxygen vacancies can ionize at relatively low temperatures. This may be written:

$$V_O \Leftrightarrow V_O^{\bullet} + e^- \qquad (3a)$$

$$V_O^{\bullet} \Leftrightarrow V_O^{\bullet\bullet} + e^- \qquad (3b)$$

An oxygen vacancy is a donor centre and three energy levels $V_O, V_O^{\bullet},$ and $V_O^{\bullet\bullet}$ are associated with such a centre. Moreover, electrons bounded with an oxygen vacancy can be transfered on $Nb^{5+}$ ion. This may be expressed as follows:

$$V_O + Nb_{Nb} \Leftrightarrow V_O^{\bullet} + Nb'_{Nb} \qquad (4a)$$

$$\text{that is} \quad V_O + Nb_{Nb} \Leftrightarrow V_O^{\bullet} + Nb^{4+}_{Nb} \qquad (4b)$$

In this way an electrically neutral complex $\left(V_O^{\bullet} Nb'_{Nb}\right)$ is created and it is a donor centre too. The charge neutrality condition may be written:

$$\left[Nb'_{Nb}\right] + \left[e^-\right] = \left[V_O^{\bullet}\right] + \left[V_O^{\bullet\bullet}\right] \qquad (5)$$

A reduced $KNbO_3$ crystal should be an n–type semiconductor at low temperatures.

According to reference data [1], a reduced $KNbO_3$ crystal is an n–type semiconductor. Donor centres are oxygen vacancies $V_O, V_O^{\bullet}$ and a complex $\left(V_O^{\bullet} Nb'_{Nb}\right)$ as well. Predominant at high temperatures double–ionized oxygen vacancies ($V_O^{\bullet\bullet}$) can be electron traps or recombination centres.

### 3.2 KNbO₃ crystals doped with Fe

Simplified synthesis reactions at the presence of $Fe_2O_3$ dopant may be expressed as follows:

$$K_2CO_3 + Nb_2O_5 \Rightarrow 2KNbO_3 + CO_2 \uparrow \qquad (6a)$$

$$K_2CO_3 + Fe_2O_3 \Rightarrow 2KFeO_2 + CO_2 \uparrow \qquad (6b)$$

$$(1-x)KNbO_3 + xKFeO_2 \Rightarrow KNb_{1-x}Fe_xO_{3-x}\left(xV_O^{\bullet\bullet}\right) \qquad (6c)$$

The occurrence of Fe ions in the Nb sublattice leads to the creation of oxygen vacancies. This may be written as follows:

$$KNb_{3-x}Fe_xO_{3-x} \equiv K^{1+}Nb^{5+}_{1-x}Fe^{3+}_xO^{2-}_{3-x}\left(xV_O^{\bullet\bullet}\right) \qquad (7)$$

The charge compensation is brought about by the creation of the double–ionized oxygen vacancy $V_O^{\bullet\bullet}$. The impurity ion and the vacancy create the pair–complex $(Fe_{Nb}'' V_O^{\bullet\bullet})$, which is an electrically neutral acceptor centre. The ionization of such an acceptor may be expressed as follows:

$$(Fe_{Nb}'' V_O^{\bullet\bullet}) \Leftrightarrow (Fe_{Nb}''' V_O^{\bullet\bullet}) + h^\bullet \qquad (8a)$$

$$\text{or} \qquad (Fe_{Nb}'' V_O^{\bullet\bullet}) \Leftrightarrow (Fe_{Nb}'' V_O^{\bullet}) + h^\bullet \qquad (8b)$$

These reactions correspond to oxygen ions valence change:

$$O_O \Leftrightarrow O_O^\bullet + e^- \quad \text{or} \quad O_O + h^\bullet \Leftrightarrow O_O \qquad (9)$$

The charge neutrality condition may be written:

$$[V_O^{\bullet\bullet}] + [V_O^\bullet] + [h^\bullet] = [Fe_{Nb}''] + [Fe_{Nb}'''] \qquad (10)$$

The doped crystal should be a p–type semiconductor.

If the doping is perform simultaneously with the reduction process the chemical reaction may be written:

$$(1-x)KNbO_{3-y} + xKFeO_2 \Rightarrow KNb_{1-x}Fe_xO_{3-x-y} \qquad (11)$$

The composition of this crystal (sample of D–type) may be expressed as follows:

$$KNb_{1-x}Fe_xO_{3-x-y} \equiv K^{1+}Nb_{1-x}^{5+}Fe_x^{3+}O_{3-x-y}^{2-}(xV_O^{\bullet\bullet})(yV_O) \qquad (12)$$

If $(Fe_{Nb}'' V_O^{\bullet\bullet})$ acceptor centres and $V_O$, $(V_O^\bullet Nb_{Nb}')$, $V_O^\bullet$ donor centres appear in crystal then the population of acceptor energy levels by electrons, which occupied previously donor levels, is possible. In such a case a compensation of the n–type conductivity follows. This may be written:

$$V_O + (Fe_{Nb}'' V_O^{\bullet\bullet}) \Leftrightarrow V_O^{\bullet\bullet} + (Fe_{Nb}''' V_O^\bullet) \,, \qquad (13)$$

with the charge neutrality condition:

$$2[V_O^{\bullet\bullet}] = [(Fe_{Nb}''' V_O^\bullet)] \qquad (14)$$

Thus Fe–doped and reduced crystal should demonstrate properties of a compensated semiconductor. When the concentration of defects is large enough impurity bands and tails of the density of states can be created [9]. The results presented in the paper [1] show clearly the occurrence of the electrical conduction in the impurity band in strongly reduced $KNbO_3$ single crystals.

The results presented in this work are in accordance with above discussion. Widths of energy gaps depend on concentration of defects. The increase of defect concentration diminishes the width of the energy gap.

The decrease in the energy gap may be connected either with a change in lattice parameters or with impurity bands in crystals with a high concentration of defects. However, optical electron transitions with the participation of tails of the density of states were observed in iron doped crystals D only. The minimum in optical transmission observed for reduced crystals B and C occurs for the photon energy $E_F$ = 1.10 eV. Reduced crystals $KNbO_3$ exhibit the n–type conductivity [10]. For a high degree of the reduction process the donor impurity band is created. Donors do not undergo a complete ionization up to about $300^0$C [1]. The impurity optical absorption band appeared in reduced crystals B and C (see Fig.1) is therefore connected with electron transitions from energy levels of the donor band to the conduction band. Acceptor energy levels connected with $(Fe_{Nb}^{''} V_O^{\bullet\bullet})$ centres appear in the crystals doped with iron. If the concentration of these centres is high enough the acceptor impurity band has been created. The presence of acceptor centres leads to the compensation of an n–type conductance connected with the nonstoichiometry. As a result of the compensation acceptor centres will be filled and donor levels connected with $V_O, V_O^{\bullet}$ and $(V_O^{\bullet} Nb_{Nb}^{'})$ centres remain empty. Electron transitions $E_V \to V_O$ and $E_V \to V_O^{\bullet}$ appear in crystals instead of transitions $V_O \to E_C$ and $V_O^{\bullet} \to E_C$. Transitions between the top of the valence band and empty donor centres diminish a value of optical transmission near the absorption edge. According to [11] the concentration of impurity of about 0.2 at.% cause the creation of the impurity band. It is very probably that in obtained reduced and Fe–doped crystals D tails of the density of states appear both below the bottom of the conduction band and the top of the valence band. The exponential absorption edge, observed in these crystals confirms this fact.

Lattice defects existing in $KNbO_3$ crystals influence on their dielectric properties (see Fig. 3). High values of dielectric permittivity, its dispersion, and dependence tan$\delta$ = f(*T*) (for low frequency) seem to point on the ionic and electron space charge polarization connected probably with surface layers and extended defects.

## 4. Summary

The obtaining of the high quality $KNbO_3$ single crystals is difficult both due to a complicated phase diagram of the solution $K_2O–Nb_2O_5$ and due to reduction processes taking place during the growth of crystals. Using the $TiO_2$ as oxidizer during the growing process enable to obtain transparent and colourless crystals.

The absorption edge is determined by direct and indirect transitions of electrons. The exponential absorption edge was observed for the crystal doped with iron. The reduction process causes the blue colouration of the crystals. The impurity optical absorption band is associated with electron transitions between the donor impurity band and the conduction band. The donor impurity band is associated with the oxygen deficiency. The compensation of the n–type conductance by the acceptor Fe dopant cause the vanishing of the impurity band and decolourization of the crystal.